\documentclass{article}
\usepackage{epsfig}

\def\gtrless{\raise2.5pt\hbox{$>$}\llap{\lower2.5pt\hbox{$<$}}}
\def\gtrapprox{\,\raise2.5pt\hbox{$>$}\llap{\lower2.5pt\hbox{$\sim$}}\,}
\newcommand{\bsq}[1]{\begin{subequations}\label{#1}}
\newcommand{\esq}{\end{subequations}}
\newcommand{\beq}{\begin{equation}}
\newcommand{\eeq}{\end{equation}}
\newcommand{\beqa}[1]{\begin{eqnarray}\label{#1}}
\newcommand{\eeqa}{\end{eqnarray}}

\def\lessapprox{\raise2.5pt\hbox{$<$}\llap{\lower2.5pt\hbox{$\approx$}}}

\begin{document}

\title{Arrest and Flow of Colloidal Glasses\footnote{Invited Plenary Talk, Th2002, Paris. To appear in Annales Henri Poincar\'e, 2003}} 
 
\author{M. E. Cates\\ 
School of Physics, The University of
Edinburgh,\\ JCMB King's Buildings, Edinburgh EH9 3JZ, GB}  

\maketitle 
\begin{abstract}
I review recent progress in understanding the arrest and flow behaviour of colloidal glasses, based on mode coupling theory (MCT) and related approaches.
MCT has had notable recent successes in predicting the re-entrant arrest behaviour of colloids with short range attractions. Developments based upon it offer important steps towards calculating, from rational foundations in statistical mechanics, nonlinear flow parameters such as the yield stress of a colloidal glass. An important open question is why MCT works so well. 
\end{abstract}

\section{Introduction}
\subsection{Soft Matter}
This paper addresses issues of arrest and flow in soft condensed matter; see \cite{book} for a useful ensemble of background reading and \cite{larson} for a good, experimentally motivated overview. We consider a system of $N$ particles in the size range of nanometres (e.g. globular proteins) to microns (traditional colloids), suspended in a solvent, with total volume $V$. To a good enough approximation, at least for equilibrium properties, the solvent degrees of freedom can be integrated out to give an effective pairwise Hamiltonian $H =\sum_{i>j}u(r_{ij})$ (in an obvious notation). Equilibrium statistical mechanics, when applicable, is governed by the partition function $Z = \int \exp(-\beta H){\cal D}[r_i]$; quantum effects play no role \cite{frenkel}.

In many colloidal materials the effective interaction $u(r)$ comprises a hard core repulsion, operative at separation $2a$ with $a$ the particle radius, combined with an attraction at larger distance. For simplicity one can imagine a square well potential of depth $\epsilon$ and range $\xi a$, with $\xi < 1$. Unlike atomic systems, to which colloidal ones are otherwise quite analagous, these parameters can be varied easily in experiment, essentially by varying the solvent conditions \cite{frenkel}. For example, adding polymers to a colloidal system will mediate an entropic attraction between spheres whose range is comparable to the size of the polymer coils and whose depth is controlled by their concentration. For globular proteins the same tricks can be played with salt concentration and pH. 

The resulting equilibrium phase diagrams are well known, and depend on the range parameter $\xi$ and the attraction energy or temperature through the parameter $\beta \epsilon$. For $\xi\gtrapprox 0.2$, at small $\beta\epsilon$ there is a phase separation from a colloidal fluid to a colloidal crystal. At higher $\beta\epsilon$, a liquid phase intervenes; the fluid undergoes a gas-liquid separation at intermediate densities although the crystal is stable at higher ones. So far, this is just like the phase diagram of argon or a similarly classical atomic substance. (But of course solvent fills the space between the colloids, so the gas is not a real gas.) However, for smaller $\xi$ the liquid phase is missing: one has only one transition, from fluid to crystal. In fact, though, the liquid phase is still lurking beneath: it is metastable. 

\subsection{Arrest in Colloidal Fluids}
Colloidal fluids can be studied relatively easily by light scattering \cite{pine}. This allows one to measure the dynamic structure factor $S(q,t) = \langle\rho({\bf q},t')\rho({\bf -q},t'+t)\rangle/N$ and also the static one, $S(q)= S(q,0)$. Here $\rho({\bf r},t)) = \sum_i\delta({\bf r}_i(t)-{\bf r})- N/V$; this is the real space particle density (with the mean value subtracted), and $\rho({\bf q},t)$ is its Fourier transform. For particles with hard-core repulsions, $S(q)$ exhibits a peak at a value $q^*$ with $q^* a = {\cal O}(1)$. The dynamic structure factor $S(q,t)$, at any $q$, decays monotonically from $S(q)$ as $t$ increases. In an ergodic colloidal fluid, $S(q,t)$ decays to zero eventually: all particles can move, and the density fluctuations have a finite correlation time. In an arrested state, which is nonergodic, this is not true. Instead the limit $S(q,\infty)/S(q) = f(q)$ defines the {\em nonergodicity parameter}. (Note that this corresponds to the Edwards-Anderson order parameter in spin glasses.) The presence of nonzero $f(q)$ signifies frozen-in density fluctuations. Although $f(q)$ is strongly wavevector dependent, it is common to quote only $f(q^*)$ \cite{kob}.

Colloidal fluids are found to undergo nonergodicity transitions into two different broad classes of arrested nonequilibrium state. One is the colloidal glass, in which arrest is caused by the imprisonment of each particle in a cage of neighbours. This occurs even for $\epsilon = 0$ (i.e. hard spheres) at volume fractions above about $\phi \equiv 4\pi a^3N/3V \simeq 0.58$. Such a system would, in equilibrium, be a crystal; but equilibrium can be delayed indefinitely once a glass forms (particularly if there is a slight spread in particle size $a$, which helps suppress nucleation). The nonergodicity parameter for the colloidal glass obeys $f(q^*)\simeq 0.7$.  The second arrested state is called the colloidal gel. Unlike the repulsive glass, the arrest here is driven by attractive interactions, resulting in a bonded network structure. Such gels can be unambiguously found, for short range attractions, whenever $\beta\epsilon \gtrapprox 5-10$. Hence it is not necessary that the local bonds are individually irreversible (this happens, effectively, at $\beta\epsilon \gtrapprox 15-20$); and when they are not, the arrest is collective, not local. It is found experimentally that for colloidal gels, $f(q^*) \gtrapprox 0.9$, which is distinctly different from the colloidal glass. The arrest line for gel formation slices across the equilibrium phase diagram (e.g., plotted on the $(\phi,\beta\epsilon)$ plane), and, depending on $\xi$, parts of it lie within two phase regions. This, alongside any metastable gas-liquid phase boundary that is present, can lead to a lot of interesting kinetics \cite{poon,kroy}, in which various combinations of phase separation and gelation lead to complex microstructures and time evolutions.

\subsection{A Brief Primer on Mode Coupling Theory (MCT)}
This is not the place to explain MCT in detail. The most powerful form of the theory \cite{goetze}, which is favoured by most of the true experts, remains somewhat obscure to many others. However, in a stripped down version (see e.g. \cite{ramaswamy,kk}) the theory can be viewed as a fairly standard one-loop selfconsistent approach to an appropriate dynamical field theory. 

We take $\beta = 1$ and start from the Langevin equations $\dot {\bf r}_i = {\bf F}_i + {\bf f}_i$ for independent particles of unit diffusivity ($D_0=1$) subjected to external forces ${\bf F}_i$. The noise force then obeys $\langle {\bf f}_i{\bf f}_j\rangle = {\bf 1} \delta_{ij}$. By standard manipulations one proceeds to a Smoluchowski equation $\dot \Psi = \Omega \Psi$ for the $N$-particle distribution function $\Psi$, with evolution operator is $\Omega = \sum_i\nabla_i.(\nabla_i-{\bf F}_i)$. Now take the forces ${\bf F}_i$ to derive (via ${\bf F}_i = -\nabla_i H$) from an interaction Hamiltonian 
\beq
H = -\frac{1}{2}\int d^3{\bf r}d^3{\bf r'} \rho({\bf r})\rho({\bf r'})c(|{\bf r}-{\bf r'}|)
\label{hamiltonian}
\eeq
where $N c(q) = V[1-S(q)^{-1}]$. This is a harmonic expansion in density fluctuations; $c(q)$ is called the direct correlation function, and its form is fixed by requiring that $S(q)$ be recovered in equilibrium. Neglected are solvent mediated dynamic forces (hydrodynamic couplings); these mean that in principle the Langevin equations for the particles should have correlated noise. Also neglected are anharmonic terms in $H$; to regain the correct higher order density correlators (beyond the two point correlator $S(q)$) in equilibrium, these terms would have to be put back.

These assumptions give a Langevin equation for the density $\rho({\bf r})$:
\beq
\dot \rho = \nabla^2\rho + \nabla(\rho\nabla\delta H/\delta\rho) + \nabla.{\bf h}
\eeq
where $\bf h$ is a suitable noise (actually with a nontrivial density dependence \cite{dean}). This equation is nonlinear, even with the harmonic choice of $H$. 
However, from it one can derive a hierarchy of equations of motion for correlators such as $S(q,t)$, more conveniently expressed via $\Phi(q,t) \equiv S(q,t)/S(q)$. Factoring arbitrarily the four-point correlators that arise in this hierarchy into products of two $\Phi$'s, one obtains a closed equation of motion for the two point correlator as
\beq
\dot\Phi(q,t) + \Gamma(q)\left[ \Phi(q,t) + \int_0^t m(q,t-t')\dot\Phi(q,t)\right] = 0
\label{correlator}
\eeq
where $\Gamma(q) = q^2/S(q)$ is an initial decay rate, and the memory function obeys
\beq
m({\bf q},t)= \sum_{{\bf k}} V_{{\bf q},{\bf k}}\Phi({\bf k},t)\Phi({\bf k-q},t)
\label{memory}
\eeq
with the vertex 
\beq
V_{{\bf q},{\bf k}} = \frac{N}{2V^2q^4}S(q)S(k)S(|{\bf k-q}|)[{\bf q}.{\bf k}c(k) + {\bf q}.(
{\bf k}-{\bf q}) c(|{\bf k-q}|)]^2
\label{vertex}
\eeq

\subsection{Dynamic Bifurcation}

The MCT equations exhibit a bifurcation that corresponds to a sudden arrest transition, upon smooth variation of either the density $\phi$ or other  parameters controlling the kernel in the harmonic hamiltonian $H$. This is best seen in the nonergodicity parameters $f(q)$, which suddenly jump (for all $q$ at once) from zero to nonzero values.  Near this (on the ergodic side, which is always the direction MCT approaches from), $\Phi(q,t)$ develops interesting behaviour. Viewed as a function of time,  it decays onto a plateau of height $f(q)$, stays there for a long time, and then finally decays again at very late times. The two decays are called $\beta$ and $\alpha$ respectively. Upon crossing the bifurcation, the $\alpha$ relaxation time diverges smoothly with the parameters; upon crossing the locus of this divergence, $f(q) \equiv S(q,\infty)$ therefore jumps discontinuously from zero to a value that is finite for all $q$. 

\section{Critiques and Defence of MCT}

This mathematical structure means that $S(q)$ fixes all of the dynamics (up to a scale factor which was the bare diffusion constant of one particle, here set to unity). As a result, the theory has significant predictive power: MCT gives, in terms of $S(q)$, the critical parameter values where arrest occurs; the power law exponents governing the $\alpha$ and $\beta$ decays; the divergence of the $\alpha$ relaxation time; and $f(q)$. 
This makes the theory falsifiable. Indeed, it has successfully been falsified. For example, there is no doubt that MCT gives the wrong density $\phi_g$ for the glass transition in hard spheres. On the other hand, the predictions (including that for $\phi_g$) do all agree with experiments at about the 10 percent level, and often better. 

Attitudes among theorists to MCT for colloids therefore vary considerably. Some argue that the approximations made are uncontrolled (true) and that, even though there are no explicit adjustable parameters in the theory, its approximations have been implicitly tuned to suit the problem at hand (normally taken to be the glass transition in hard spheres). These might be fair accusations in part, but if so they can also be levelled at much of {\em equilibrium} liquid state theory, where, for decades, ad-hoc closures (Percus-Yevick, hypernetted chain, etc.) have competed for survival by an essentially Darwinian process. Other theorists point to the already-falsified status of MCT, drawing attention to the misplaced $\phi_g$ or, more interestingly, to other physical situations where exactly the same kind of approximation leads to totally wrong predictions (e.g. spurious arrest in systems that are known to evolve smoothly towards a unique equilibrium state). 

Against all this must be weighed the continuous stream of experimental results, many of them subtle but others not, that confirm the MCT predictions in surprising detail --- especially if one makes one or two ad-hoc `corrections' to adjust parameters like $\phi_g$ \cite{goetze2}. (There are also many simulation results which increasingly confirm the same picture \cite{kob}.) 

\begin{figure}[h]
\centerline{\psfig{file=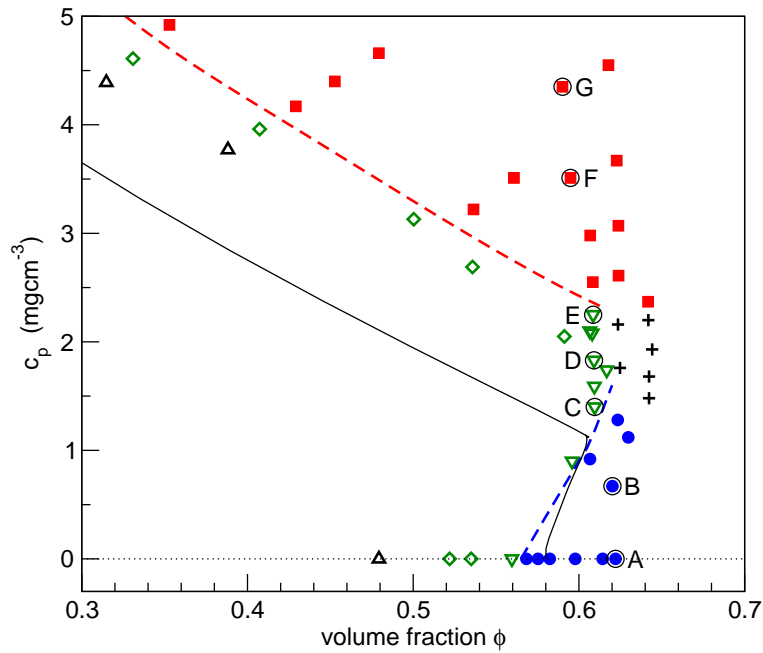,width=10.cm}}  
\caption{Experimental test of MCT for colloids with short range attractions. The dashed line separates ergodic samples (open symbols) from nonergodic samples, identified as gels (squares), glasses (circles) and intermediate ($+$). The control parameter on the vertical axis is the concentration of added polymer; horizontal is the volume fraction. The solid line is the MCT prediction for the nonergodicity transition (shifted to give the correct $\phi_g$ in the absence of polymer). Courtesy K. Pham; see Poon et. al. \protect\cite{pham1} for individual discussions of the samples marked A-G.} 
\label{fig1}
\end{figure}

\begin{figure}[h]
\centerline{\psfig{file=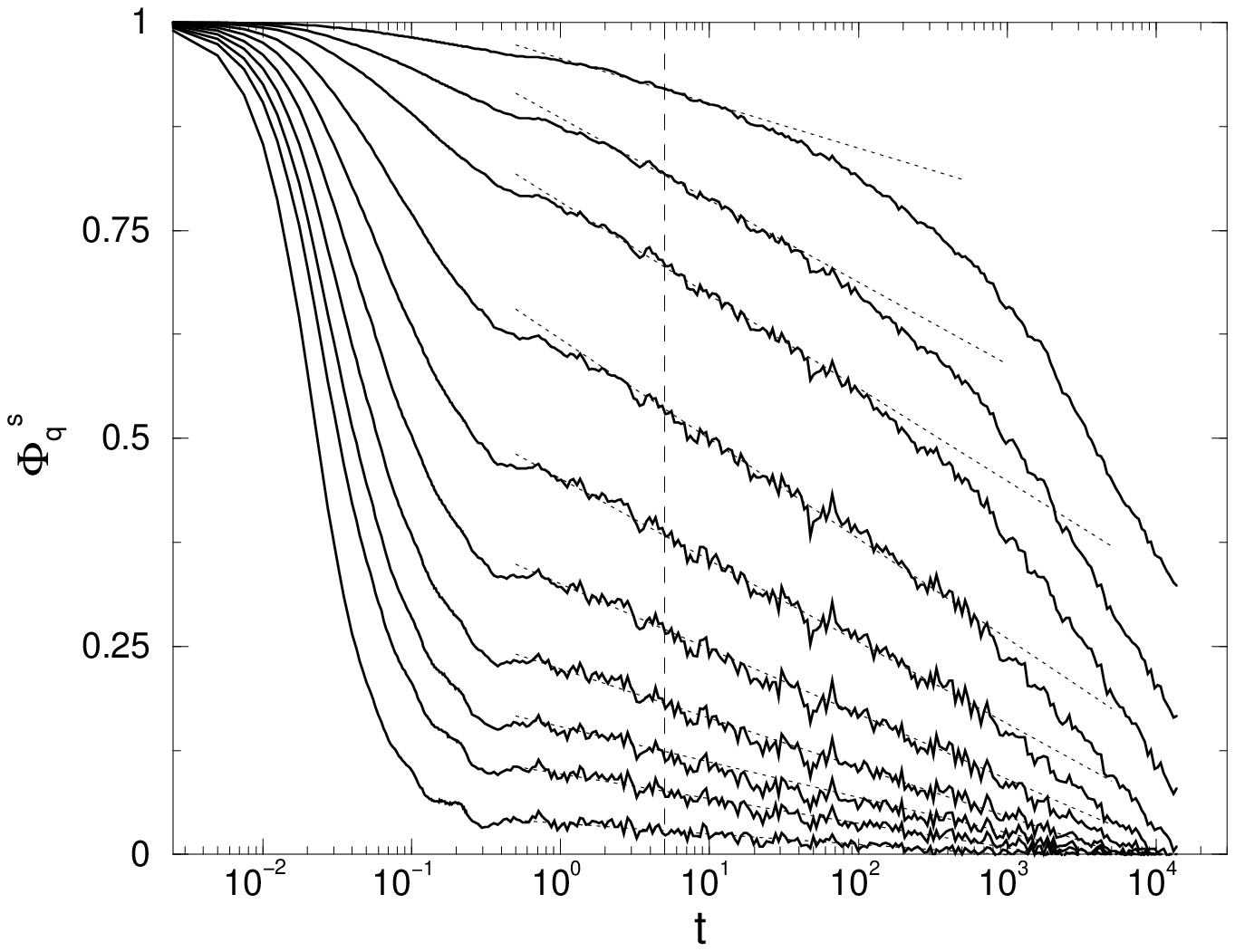,width=10.cm}}  
\caption{Simulation test of MCT for Brownian spheres with short range attractions. The correlator $\Phi(q,t)$ is shown for various $q$, in a system thought to lie close to the higher order singularity at the glass-gel corner. Dotted lines are fits to the logarithmic time dependence predicted by MCT in this region. (From \protect\cite{puertas}.)} 
\label{fig2}
\end{figure}

A striking recent success of MCT concerns systems with attractive interactions as well as hard-core repulsions, in a region of parameter space where not one but two arrest transitions are nearby (these are the transitions to a colloidal glass and to a colloidal gel). First, MCT unambiguously predicts \cite{dawson} that adding a weak, short range attraction to the hard sphere system should melt the glass. The mechanism appears to be a weakening of the cage of particles in which a given particle is trapped. (This is caused by members of that cage moving closer together under the attractive forces; gaps in the cage, allowing motion, are then more likely to appear.) Second, MCT predicts that adding more of the same attraction should mediate a second arrest, this time into a gel ($f(q^*)\sim 0.95$). Third, MCT predicts that as parameters are varied, a higher order bifurcation point should be seen, when the re-entrant arrest line (implied by the above picture) crosses from being a smooth curve to being a cuspy one \cite{dawson}. Although not every detail of this scenario is yet confirmed, there is clear experimental evidence of the predicted re-entrant behaviour connecting the glass and the gel arrest lines \cite{pham1} (see Figure \ref{fig1}) and clear evidence of the proximity of the higher order bifurcation, which shows up as a characteristic logarithmic decay for $\Phi(q,t)$ \cite{pham3}. The latter is also seen clearly in recent simulations \cite{puertas}; see Figure \ref{fig2}.

It is worth emphasizing that these results were predicted by MCT before the experiments (or simulations) were actually begun. The success of MCT at unifying the glass and gel arrest transitions in attractive colloids does much to dispel fears of  implicit parameter tuning during the earlier evolution of the theory. It reconfirms the {\em generic} success of the MCT approach, at around the 10 percent level, in describing interacting colloids. 

So, perhaps the time has come to stop criticising MCT for colloids on the grounds that it `cannot be right'. Indeed it cannot: the problem is really to understand how it can do so well. To paraphrase Churchill \cite{churchill}: Mode coupling theory is the {\em worst} theory of colloidal glasses -- apart from all the others that have been tried from time to time.

\section{Shear Thinning}

In Refs.\cite{fuchsprl}, M. Fuchs and the author developed a theory, along MCT lines, of colloidal suspensions under flow. The work was intended mainly to address the case of repulsion-driven glasses, and to study the effect of imposed shear flow either on a glass, or on a fluid phase very near the glass transition. In either case, simplifications might be expected because the bare diffusion time $\tau_0 = a^2/D_0$ is small compared to the `renormalized' one $\tau = a^2/D$, which in fact diverges (essentially as the $\alpha$ relaxation time) as the glass transition is approached. If the imposed shear rate (which we assume steady) is $\dot\gamma$, then for $\dot\gamma\tau_0 \ll 1 \leq \dot\gamma\tau$, one can hope that the details of the local dynamics are inessential and that universal features related to glass formation, should dominate. Note, however, that by continuing to use a quadratic $H$ (Eq.\ref{hamiltonian}), we will assume that, even under shear, the system remains `close to equilibrium' in the sense that the density fluctuations that build up remain small enough for a harmonic approximation to be useful.

The basic route followed in Ref.\cite{fuchsprl} is quite similar to that already laid out above for standard MCT. However, we assume that an imposed shear flow is present; obviously this changes the equations of motion. A key simplification is to neglect velocity fluctuations so that the imposed shear flow is locally identical to the macroscopic one; this cannot be completely correct, but allows progress to be made. For related earlier work see Refs.\cite{indrani,milner}. 

We again take $\beta = 1$, $D_0 = 1$, and start from the Langevin equations $\dot {\bf r}_i = {\bf u} + {\bf F}_i + {\bf f}_i$ for independent particles of unit diffusivity subjected to external forces ${\bf F}_i$ and, now, an imposed flow velocity ${\bf u}({\bf r}_i)$. We take this to be a simple shear flow with ${\bf u}({\bf r}) = \dot\gamma y {\bf \hat x}$. The Smoluchowski equation $\dot \Psi = \Omega \Psi$ is unchanged but the evolution operator is now $\Omega = \sum_i\nabla_i.(\nabla_i-{\bf F}_i-{\bf u}({\bf r}_i))$. We again take the forces ${\bf F}_i$ to derive from Eq.\ref{hamiltonian}, and with the same assumptions as before gain
a Langevin equation for the density $\rho({\bf r})$:
\beq
\dot \rho +{\bf u}.\nabla\rho=  \nabla^2\rho + \nabla(\rho\nabla\delta H/\delta\rho) + \nabla.{\bf h}
\eeq

So far, the adaption to the equations to deal with imposed shearing is fairly trivial. The next stages are not. We assume an initial equilibrium state with $\Psi(t=0) \propto Z$, and switch on shearing at $t=0+$. We define an {\em advected correlator}
\beq
\Phi({\bf q},t) = \langle \rho({\bf q},0)\rho(-{\bf q}(t),t)\rangle/S(q)N
\eeq
where ${\bf q}(t) = {\bf q} + {\bf q}.{\bf K} t$ with ${\bf K}$ the velocity gradient tensor, $K_{ij} = \dot\gamma\delta_{ix}\delta_{jy}$. This definition of the correlator subtracts out the trivial part of the advection, which is merely to transport density fluctuations from place to place. The nontrivial part comes from the effect of this transport on their time evolution; the main effect (see e.g. \cite{milner}) is to kill off fluctuations by moving their wavenumbers away from $q^*$ where restoring forces are weakest (hence the peak there in $S(q)$). Hence the fluctuations feel a stronger restoring force coming from $H$, and decay away more strongly. This feeds back, through the nonlinear term, onto the other fluctuations, including ones transverse to the flow and its gradient (i.e., with ${\bf q}$ along $z$) for which the trivial advection is absent.

There follow a series of MCT-like manipulations which differ from those of the standard approach because they explicitly deal with the switchon of the flow at $t=0+$ and integrate through the transient response to obtain the steady state correlators, under shear, as $t\to\infty$. There is no integration through transients in standard MCT; on works directly with steady-state quantities. (In practice also, the following results were obtained in Refs. \cite{fuchsprl} using a projection operator formalism which differs in detail from the version of MCT outlined above.) Despite all this, the structure of the resulting equations is remarkably similar to Eqs. \ref{correlator},\ref{memory}:
\beq
\dot\Phi({\bf q},t) + \Gamma({\bf q},t)\left[\Phi({\bf q},t)+\int_0^tm({\bf q},t,t')\dot\Phi({\bf q},t')\right] = 0
\eeq
with $\Gamma(q)$ replaced by a time dependent, anisotropic quantity:
\beq
\Gamma({\bf q},t)S(q) = q^2 + q_xq_y\dot\gamma t+ (q_xq_y\dot\gamma t+ q_x^2\dot\gamma^2 t^2)S(q) - q_xq_yS(q)/q
\eeq
The memory kernel is no longer a function of the time interval $t-t'$ but depends on both arguments separately
\beq
m({\bf q},t,t')= \sum_{{\bf k}} V({\bf q},{\bf k},t,t')\Phi({\bf k},t)\Phi({\bf k-q},t)
\eeq
through a time-dependent vertex $V$, too long to write down here \cite{fuchsprl}.

Using a nonequilibrium Kubo-type relationship, one can also obtain an expression for the steady state viscosity $\eta = \sigma(\dot\gamma)/\dot\gamma$ where $\sigma(\dot\gamma)$ is the shear stress as a function of shear rate. The viscosity is expressed as an integral of the form
\beq
\eta = \int_0^\infty dt \sum_{{\bf k}} F({\bf k},t) \Phi^2({\bf k},t)
\label{kubolike}
\eeq
where the function $F$ may be found in Ref. \cite{fuchsprl}.

\subsection{Results}
The above calculations give several interesting results. 
First, any nonzero shear rate, however small, restores ergodicity for all wavevectors (including ones which are transverse to the flow and do not undergo direct advection). This is important, since it is the absence of ergodicity that normally prevents MCT-like theories being used inside the glass phase, at $T<T_g$ or $\phi>\phi_g$. Here we may use the theory in that region, so long as the shear rate is finite.

In the liquid phase ($\phi<\phi_g$) the resulting flow curve $\sigma(\dot\gamma)$ shows shear thinning at $\dot\gamma\tau \gtrapprox 1$, that is, when the shearing becomes significant on the timescale of the slow relaxations. This is basically as expected. Less obviously, throughout the glass, one finds that the limit $\sigma(\dot\gamma \to 0+) \equiv \sigma_Y$ is nonzero. This quantity is called the yield stress and represents the minimum stress that needs to be applied before the system will respond with a steady-state flow. (For lower stresses, various forms of creep are possible, but the flow rate vanishes in steady state.) 

The prediction of a yield stress in colloidal glasses is significant, because glasses, operationally speaking, are normally defined by the divergence of the viscosity. However, it is quite possible for the viscosity to diverge without there being a yield stress, for example in `power law fluids' where $\sigma(\dot\gamma) \sim \dot\gamma^p$ with $0<p<1$. Indeed, a recent model of `soft glassy materials' (designed for foams, emulsions, etc., and based on the trap model of Bouchaud \cite{bouchaud})  gives a well-developed power law fluid region above $T_g$ in which the viscosity is infinite but the static shear modulus zero \cite{sgr}. 
This does not happen in the present calculation, where the yield stress jumps discontinuously from zero to a nonzero value, $\sigma_Y^c$, at $\phi_g$. The existence of a yield stress seems to be in line with most experimental data on the flow of colloidal glasses, although one must warn that operational definitions of the yield stress do vary across the literature \cite{barnes}. Ours is defined as the limiting stress achieved in a sequence of experiments at ever decreasing $\dot\gamma$, ensuring that {\em a steady state is reached} for each shear rate before moving onto the next one. The latter requirement may not be practically achievable since the equilibration time could diverge (certainly one would expect to have to wait at least for times $t$ such that $\dot\gamma t \gtrapprox 1$). But unless the flow curve has unexpected structure at small shear rates, the required extrapolation can presumably be made. 

\begin{figure}[h]
\centerline{\psfig{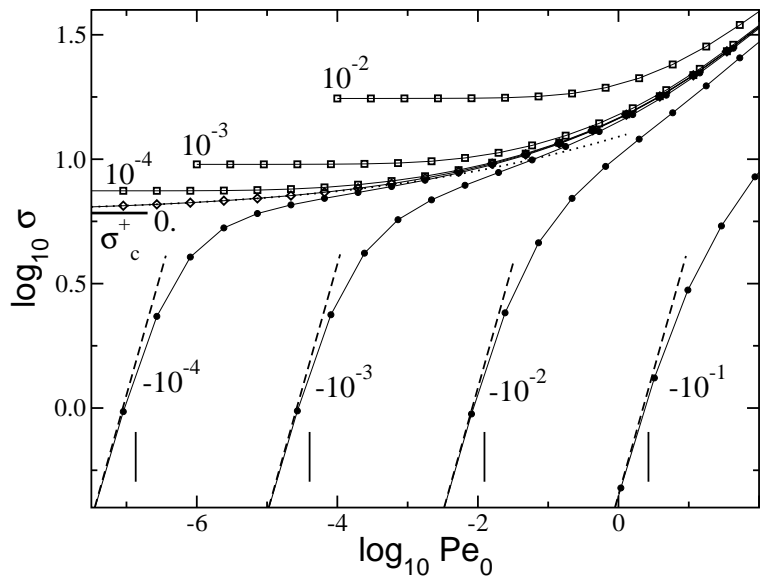}}  
\caption{Flow curves for the ISHSM. The shear stress is given in units of $kT/d^3$ with $d$ the particle diameter. The parameter Pe$_0$ is $\dot\gamma d^2/D_0$ with $D_0$ the bare diffusion constant. The curves marked with circles are within the fluid phase (with dashed asymptotes showing the Newtonian limit); the curves marked with squares are in the glass. The diamonds denote the critical case. Each curved is marked by its distance from the glass transition, $\phi-\phi_g$. The yield stress at the glass transition $\sigma_Y^c$ (here called $\sigma_c^+$) is indicated on the left.} 
\label{fig3}
\end{figure}

The existence of a yield stress everywhere within the glass phase follows from the structure of the MCT-inspired calculations outlined above and detailed in Ref. \cite{fuchsprl}. However, to calculate an actual value for $\sigma_Y$ requires further approximations; these avenues are pursued in Ref. \cite{fuchsprl}. Quantitative results for one such approximation, called the `isotropically sheared hard sphere model' (ISHSM), are given in Figure \ref{fig3}. Such approximations can also give values for the flow curve exponent after the onset of yield ($\sigma-\sigma_Y\sim\dot\gamma^p$ with $p \simeq 0.15$), and for the growth of the yield stress beyond the glass transition ($\sigma_Y - \sigma_Y^c \sim (\phi-\phi_g)^{1/2}$).

\subsection{Schematic MCT models}
It has long been known that the key mathematical structure behind (conventional, unsheared) MCT can be captured by low-dimensional schematic models in which the full ${\bf q}$ dependence is suppressed. In other words, one chooses a single mode, with a representative wavevector around the peak of the static structure factor, and writes mode coupling equations for this mode in isolation. At a phenomenological level, one can capture the physics similarly even with shearing present (despite the more complicated vectorial structure that in reality this implies). Specifically one can define the $F_{12}^{\dot\gamma}$ model --- the sheared extension of a well known static model, $F_{12}$ --- via
\beq
\dot\Phi(t) + \Gamma \left[ \Phi(t) + \int_0^t m(t-t')\dot\Phi(t') dt'\right] = 0
\label{scheme1}\eeq
with memory function (schematically incorporating shear)
\beq
m(t) = [v_1\Phi(t)+v_2\Phi^2(t)]/(1+\dot\gamma^2t^2)
\label{scheme2}\eeq
The vertex parameters $v_{1,2}$ are smooth functions of the volume fraction $\phi$ (and any interactions).
To calculate flow curves, etc., one also needs a schematic form of Eq.\ref{kubolike}; here we take the first moment of the correlator to fix the time scale for stress relaxation (which is, in suitable units, simply the viscosity):
\beq
\eta = \int_0^\infty \Phi(t) dt
\label{scheme3}\eeq
(Note that a different choice, e.g. with $\Phi(t)^2$ in this equation to closer resemble Eq.\ref{kubolike}, would yield quite similar results.)
This schematic model gives very similar results to the ISHSM, with $\sigma-\sigma_Y\sim \dot\gamma^{0.16}$ and $\sigma_Y-\sigma_Y^c\sim(\phi-\phi_g)^{1/2}$ \cite{fuchsprl}.  The qualitative reproducibility of these results within different types of approximation scheme is reassuring.

\section{Shear Thickening and Jamming}
\begin{figure}[h]
\centerline{\psfig{file=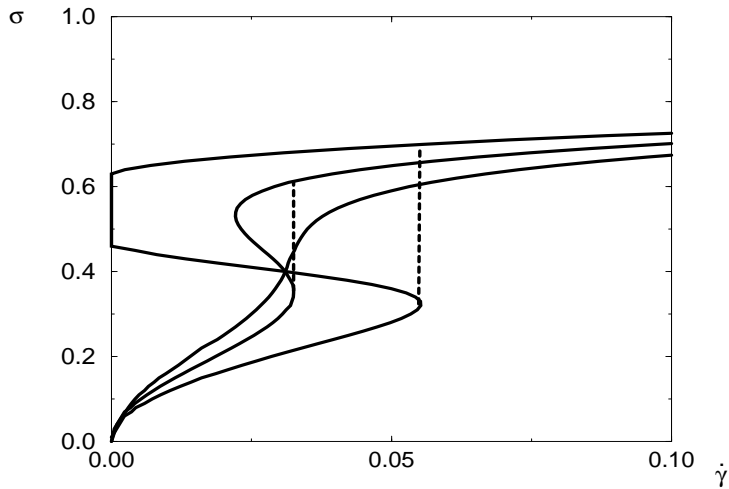,width=10.cm}}  
\caption{Three possible flow curves for a shear thickening material. The monotonic curve corresponds to continuous shear thickening. The remaining two curves are S-shaped; one expects, on increasing the shear rate, the stress to jump from the lower to upper branch at (or before) the vertical dashed line shown in each case. One curve shows the full jamming scenario: the existence of an interval of stress, here between 0.45 and 0.63, within which the flow rate is zero, even in a system ergodic at rest. (Stress and strain rate units are arbitrary.)} 
\label{fig4}
\end{figure}

The calculations described above predict, generically, shear thinning behaviour: advection kills fluctuations, reducing the $\alpha$ relaxation time, which causes the system to flow more easily at higher stresses. However, in some colloidal systems, the reverse occurs. This is shear thickening, and gives a flow curve $\sigma(\dot\gamma)$ with upward curvature. In extreme cases, an essentially vertical portion of the curve is reported \cite{discthick}. One interpretation of the latter scenario (called `discontinuous shear thickening') is that the underlying flow curve is actually S-shaped. Since any part of the curve with negative slope is mechanically unstable (a small increase in the local shear rate would cause an acceleration with positive feedback), this allows a hysteresis cycle in which, at least according to the simplest models, discontinuous vertical jumps on the curve bypass the unstable section (see Figure \ref{fig4}). 

If this viewpoint is adopted, there seems to be nothing to prevent the upper, re-entrant part of the curve to extend right back to the vertical axis (see Figure \ref{fig4}) in which case there is zero steady-state flow within a certain interval of stress. The system has both an upper and a lower yield stress delimiting this region. (If it is nonergodic at rest, it could also have a regular yield stress on the lower part of the curve near the origin -- we ignore this here.)  This case has been called `full jamming' \cite{head}. Although mostly a theoretical speculation, one or two experimental reports of this kind of behaviour have appeared in the literature recently \cite{bibette}.

The above discussion suggests that shear thickening and full jamming might be viewed as a stress-induced glass transition of some sort. If so, it is natural to ask whether this idea can be accommodated within an MCT-like approach.
Since the analysis of Ref. \cite{fuchsprl} gives only shear thinning, this is far from obvious. In particular, a stress-induced glass transition would require the vertex $V$ to `see' the stress; this might require one to go beyond harmonic order in the density, that is, it might require improvement to Eq.\ref{hamiltonian}. Indeed, since it is thought that jamming arises by the growth of chainlike arrangements of strong local compressive contacts \cite{jamming}, it is reasonable that correlators beyond second order in density should enter. 

In very recent work, an ad-hoc schematic model along the lines of Eqs.\ref{scheme1}--\ref{scheme3} has been developed to address shear thickening. The vertex is ascribed explicit dependence not only on $\dot\gamma$ (as suggested by the shear thinning calculations of Ref. \cite{fuchsprl}) but also on the shear stress $\sigma$. It is found that a vertex which is monotonically decreasing with $\dot\gamma$ but monotonically increasing with $\sigma$ can indeed result in shear thickening, and, under some conditions, in full jamming \cite{holmes}. This work is preliminary, but interesting in that it suggests how new physics (beyond two-point correlations) may need to be added to MCT before the full range of observed colloidal flow behaviour is properly described. Of course, even for systems at rest, it is known that some important physics is missing from MCT, in particular various kinds of `activated dynamics' in which the system can move exponentially slowly despite being in a region of phase space where, according to MCT, it cannot move at all (see e.g. \cite{kk}). Jamming seems very different from this: so perhaps there are more things missing from MCT than just activated processes.  

\section{Conclusion}
Mode Coupling Theory (MCT) has had important recent successes, such as predicting, in advance of experiment, the re-entrant glass/gel nonergodicity curves that arise in colloidal systems with short range attractions \cite{dawson,pham1}.

Theoretical developments directly inspired by MCT now offer a promising framework for calculating the nonlinear flow behaviour of colloidal glasses and glassy liquids \cite{fuchsprl}. In fact, this is the only {\em quantitative} framework currently in prospect for the rational prediction of yield behaviour and nonlinear rheology in this or any other class of nonergodic soft materials. (Other work on the rheology of glasses \cite{sgr,bbk} does not, as yet, offer quantitative prediction of experimental quantities.) While promising, many things are missing from our approach: velocity fluctuations, hydrodynamic forces, anharmonicity in $H$ (possibly implicated in shear-thickening) etc., are all ignored. Further hard theory work is needed here.

The eventual goal lies beyond the steady state flow curve (shear stress as a function of shear rate, $\sigma(\dot\gamma)$) discussed in this article. Ideally we would like a full constitutive equation that relates the stress tensor at a given time to the preceding deformation history (or vice versa). To do this, starting from statistical mechanics, is difficult for any class of material; its achievement for the case of entangled polymers \cite{doiedwards}, which are ergodic, was a highlight of theoretical physics in the late 20th century. To obtain a well-founded constitutive equation for a significant class of nonergodic soft materials is a worthy goal for the early 21st. 

As a framework for theoretical prediction, MCT is easy to criticize, but much harder to improve. The key question that should exercise the minds of theorists is no longer `is it correct' (it clearly is not, in a technical sense) but `why does it work so well'? Only by engaging with this question are we are likely to improve our understanding of nonergodic colloidal materials and, in the longer run, come up with something better.

{\em Acknowledgements:} I am deeply indebted to Matthias Fuchs, much of whose work is reviewed above, for introducing me to MCT.

\end{document}